\documentclass[review,12pt,a4paper,3p]{elsarticle}
\usepackage[version=3]{mhchem} 
\usepackage[]{geometry}
\usepackage[scriptsize,tight]{subfigure}
\usepackage{graphicx}
\usepackage{multirow}
\usepackage{pdflscape}
\usepackage{dcolumn}
\usepackage{bm}
\usepackage{color,soul}
\usepackage{hyperref}

\begin{document}

\begin{frontmatter}

\title{Targeting Receptor Binding Domain and Cryptic Pocket of Spike glycoprotein from  SARS-CoV-2 by biomolecular modeling}

\author{Kewin Otazu$^{1}$}

\author{Manuel E. Chenet-Zuta$^{2}$}

\author{Georcki Rop\'on-Palacios$^{1}$\corref{coric}}
\ead{georopon@gmail.com}

\author{Gustavo E. Olivos-Ram\'{\i}rez$^{3}$}

\author{Gabriel M. Jimenez-Avalos$^{3}$}

\author{Cleidy Osorio-Mogoll\'on$^{4}$}

\author{Frida Sosa-Amay$^{5}$}

\author{Rosa Vargas-Rodr\'{\i}guez$^{5}$}

\author{Tania P. Nina-Larico$^{1}$}

\author{Riccardo Concu$^{6}$}

\author{Ihosvany Camps$^{1}$\corref{coric}}
\ead{icamps@unifal-mg.edu.br}

\cortext[coric]{Corresponding author}

\address{$^{1}$Laborat\'orio de Modelagem Computacional - \emph{La}Model, Instituto
de Ci\^{e}ncias Exatas- ICEx. Universidade Federal de Alfenas - UNIFAL-MG,
Alfenas, Minas Gerais, Brasil}

\address{$^{2}$Escuela de Posgrado, Universidad San Ignacio de Loyola, Lima, Per\'u}

\address{$^{3}$Facultad de Ciencias y Filosof\'{\i}a, Universidad Peruana Cayetano Heredia, Lima, Per\'u}

\address{$^{4}$Faculdade de Medicina de Ribeir\~{a}o Preto, Universidade de S\~{a}o
Paulo, S\~{a}o Paulo, Brasil}

\address{$^{5}$Facultad de Farmacia y Bioqu\'{\i}mica, Universidad Nacional de la Amazon\'{\i}a Peruana, Iquitos, Per\'u}

\address{$^{6}$LAQV-REQUINTE of Chemistry and Biochemistry, Faculty of Sciences, University of Porto, Porto, Portugal}

\begin{abstract}
SARS-CoV-2, the causative agent of the disease known as Covid-19, has so far reported around 3,435,000 cases of human infections, including more than 239,000 deaths in 187 countries, with no effective treatment currently available. For this reason, it is necessary to explore new approaches for the development of a drug capable of inhibiting the entry of the virus into the host cell. Therefore, this work includes the exploration of potential inhibitory compounds for the Spike protein of SARS-CoV-2 (PDB ID: 6VSB), which were obtained from The Patogen Box. Later, they were filtered through virtual screening and molecular docking techniques, thus obtaining a top of 1000 compounds, which were used against a binding site located in the Receptor Binding Domain (RBD) and a cryptic site located in the N-Terminal Domain (NTD), resulting in good pharmaceutical targets for the blocking the infection. From the top 1000, the best compound (TCMDC-124223) was selected for the binding site. It interacts with specific residues that intervene in the recognition and subsequent entry into the host cell, resulting in a more favorable binding free energy in comparison to the control compounds (Hesperidine and Emodine). In the same way, the compound TCMDC-133766 was selected for the cryptic site. These identified compounds are potential inhibitors that can be used for the development of new drugs that allow effective treatment for the disease.
\end{abstract}


\end{frontmatter}

\section{Introduction}
\label{Sec:Intro}
Coronaviruses (CoV) are a large family of enveloped positive chain RNA viruses, which belong to the Nidoviral order and are primarily responsible for upper respiratory and digestive tract infections, both in domestic animals and in humans~\cite{Cruz-RevistaClinicaEspanola-2020,Lopez-Nefrologia-2020}. In the last 20 years, multiples CoVs have been transmitted from animal hosts to humans, causing severe outbreaks of respiratory diseases and coming to be considered as pandemic on two occasions~\cite{Cheng-Clin.Microbiol.Rev.-20-660,Chan-Clin.Microbiol.Rev.-28-465}. Since then, multiple efforts have been made to study these viruses in different animals in order to identify potential infectious agents who affect humans~\cite{WHO-2003,Fehr_2015,WHO-2019}.

In December 2019, a new type of CoV (SARS-CoV-2), causing the disease called COVID-19, was reported in the city of Wuhan, China, which has been spreading rapidly throughout China and all the world. Currently (May 2020), the World Health Organization has reported more than 3,435,000 cases of human infections, including more than 239,000 deaths in 187 countries. That is why, since its emergence, various investigations have been carried out to understand the phylogenetic relationships of this virus, as well as the structural characteristics of its proteins, mainly those that are involved in the processes of viral replication and pathogenesis, in order to discover inhibitors thereof~\cite{Chen-Biochem.Biophys.Res.Commun.-525-135,Dong-bioRxiv-2020,WHO-2020}.

The Spike (S) protein is a multifunctional molecular machinery that mediates the entry of CoV into host cells, in addition to promoting transmission between species, especially in Betacoronaviruses ($\beta$CoV)~\cite{Gierer-J.Virol.-87-5502}. This protein belongs to class I of viral fusion proteins. That is, it needs to be cleaved by a cellular protease to bind to the host receptor~\cite{Matsuyama-J.Virol.-84-12658,Glowacka-J.Virol.-85-4122,Shulla-J.Virol.-85-873,Kawase-J.Virol.-93-19,Letko-NatureMicrobiology-5-562}.

The S protein is a $180\,kDa$ homotrimer consisting of an extracellular domain, an intermembrane domain, and an intracellular domain~\cite{Wrapp-Science-367-1260}. The extracellular domain contains the $S1$ and $S2$ subunits, related to membrane recognition and fusion, respectively~\cite{Li-Annu.Rev.Virol.-3-237}. The $S1$ subunit contains a region called Receptor Binding Domain (RBD) that mediates the recognition of the host cell~\cite{Song-PLOSPathogens-14-1007236,Letko-NatureMicrobiology-5-562}, which varies between the different coronavirus lineages, being the converting enzyme of angiotensin-2 (ACE2) the human receptor for SARS-CoV-2~\cite{Monteil-Cell-181-7}.

On the other hand, the S protein is cleaved in the inter domain of the $S1$ and $S2$ subunits, through the activity of certain host proteases, which allows the fusion of protein S with the cell membrane for the entry of the virus~\cite{Coutard-AntiviralRes.-176-104742}. These proteases are a determining factor in the host range pathogenicity, tissue tropism, and transmissibility~\cite{Millet-VirusRes.-202-120}. Protein S can be cleaved with one or more proteases, such as furin, trypsin, cathepsin, transmembrane serine protease-2 (TMPRSS-2) and Human Airway Tripsin-like Protease (HAT), which determine the mode of entry of the virus, either through the membrane or by endocytosis~\cite{Bertram-J.Virol.-85-13363}.

The structure of the SARS-CoV-2 RBD consists mainly of 5 antiparallel $\beta$ sheets ($\beta 1$, $\beta 2$, $\beta 3$, $\beta 4$ and $\beta 7$), in addition to short $\alpha$ helices connected by loops that form the core of the structure~\cite{Lan-Nature-581-215}. Between sheets $\beta 7$ and $\beta 4$ there is an extended insert that contains short $\beta$ sheets $5$ and $6$, $\alpha 4$ and $\alpha 5$ helices, and loops, which make up the recognition Binding Motif (RBM), which contains the main recognition residues~\cite{Yang-IntegrMed.Res.-9-100426}. In addition, this region has four disulfide bridges, of which $3$ of them (C336-C361, C379-C432 and C391-C525) help stabilize the core, while another disulfide bridge (C480-C488) stabilizes the loop in the final distal part~\cite{Wang-Cell-181-9}.

Several studies have described, at structural level, the way in which the RBD domain joins the ACE2 receptor peptidase (PD) domain~\cite{Li-Science-309-1864,Yang-IntegrMed.Res.-9-100426}, which is caused by rearrangement of the surface of these two molecules in the binding regions that present a highly hydrophobic environment~\cite{Li-Science-309-1864}. During RBD-ACE2 interaction, the formation of $13$ hydrogen bridges and two salt bridges occur, where residues K417, G446, Y449 N467, N487, Y489, Q493, T500, N501 G502, and Y505, located on the surface of the RBM, mediate these interactions~\cite{Lan-Nature-581-215}. In addition, in SARS-CoV-2, unlike SARS-CoV, voluminous residues (V483 and E484) located in the loop of the junction bridge have been identified, which increases the contact surface and improves recognition for the initiation of the infection~\cite{Yang-IntegrMed.Res.-9-100426}.

As proteins are dynamic structures, during their rearrangement, transient binding sites known as cryptic pocket, can originate~\cite{Vajda-Curr.Opin.Chem.Biol.-44-1,Beglov-PNAS-115-3416}. These sites are formed only when the protein structure is in its holo state (protein-ligand) and are characterized by presenting less hydrophobic residues with greater flexibility~\cite{Cimermancic-J.Mol.Biol.-428-709}. Finding these hidden sites can be a difficult task. However, in recent years, the study of these cryptic sites has taken on great importance, because they provide relevant biological information on sites not previously described in proteins that are considered pharmaceutical targets. Improve treatments in cases where the inhibition of the active site cannot occur with sufficient potency and specificity, in addition, its identification can expand the drugable protomer~\cite{Vajda-Curr.Opin.Chem.Biol.-44-1}. A clear example of a recently developed algorithm to identify cryptic sites without the need to have the structure bound to a ligand is Cryptosite~\cite{Cimermancic-J.Mol.Biol.-428-709}.

In this context, the present work addresses the drugability of the RBD domain and potential cryptic sites in the crystallographic structure of S protein. This is needed in order to identify potential inhibitors for these regions, through the use of biophysical and computational chemistry tools. Finally, to obtain an inhibitor capable of blocking the entry of the virus into the host cell.

\section{Material and methods}

\subsection{Drug database}
A library of chemical compounds was obtained from The Pathogen Box (Van Voorhis et al., 2016), frequently used in the search for treatment against malaria. This box originally contains around 20,000 compounds in SMILE format, stored in a XLSX file, which were used for virtual screening tests. The compounds were filtered based on the criteria established in Lipinsky's rule of 5~\cite{Lipinski-Adv.DrugDeliveryRev.-23-3}, selecting only those molecules that do not show any violation of the rule. Subsequently, the compounds were converted to SDF, PDB and PDBQT formats, in this order consecutively, using OpenBabel v2.4.1 software~\cite{OBoyle-J.Cheminf.-3-33}, adding polar hydrogens for pH 7.4, following the methodology described in Ref. \cite{RoponPalacios-J.Biomol.Struct.Dyn.--1-2019}. In addition, the three-dimensional structures of the compounds were minimized using the MMFF96 force field, implemented in the OpenBabel software, in order to optimize their geometry. This entire procedure was performed using an \emph{in-house} Python script that automates each of the steps described for each compound.

\subsection{Receptor preparation: binding and cryptic site}
The CryoEM-resolved structure of the SARS-CoV-2 Spike (S) protein was obtained from the Protein Data Bank~\cite{Gao-Science-368-779}in PDB format (PDB ID: 6VSB, $3.46\, \AA$ resolution)~\cite{Wrapp-Science-367-1260}. Afterwards, the structure was repaired by adding missing loops and missing atoms with SwissModel~\cite{Waterhouse-NucleicAcidsRes.-46-296}, followed by adding di-sulfide bridges (C131-C166, C291-C301, C336-C361, C379-C432, C391-C525 C538-C590, C617-C649, C662-C671, C738-C760, C743-C749, C1032-C1043 and C1082-C1126) using the PDB Reader module available from Charmm-GUI~\cite{Jo-J.Comput.Chem.-29-1859}. Subsequently, the structure was prepared with MGLTools v1.5.7~\cite{Morris-J.Comput.Chem.-30-2785}, with which Gasteiger-Marsili fillers and polar hydrogens are added, and finally it was converted to .PDBQT format.

For virtual screening assays, the RBD region, located in the $S1$ domain (RBD-$S1$) of S protein, was considered as a potential site of inhibition, since this region has been reported in various studies as the region binding to the human ACE2 receptor, mediating the onset of infection~\cite{Chen-InfectDisPoverty.-9-24,Yan-Science-367-1444,Giron-VirusRes.-285-198021,Tai-AntiviralRes.-179-104820,Monteil-Cell-181-7}. Amino acids Y396, S399 and F400 were selected as the proximal, central and distal reference point in the RBD to cover this entire domain, when configuring the simulation box both in virtual screening and in molecular docking. Likewise, putative cryptic sites were predicted, using the CryptoSite server~\cite{Cimermancic-J.Mol.Biol.-428-709}, keeping those cryptic sites that had a score of $\geq 20$. Both the RBD site and the predicted cryptic site were used for the virtual screening trials. The coordinates, and the size of the simulation box were configured on the two sites using MGLTools v1.5.7.

\subsection{Virtual screening}
With the database of the filtered chemical compounds, a virtual screening was performed on the RBD-$S1$ region and on the cryptic pocket, using the Autodock-Vina v1.1.2 software~\cite{Trott-J.Comput.Chem.-31-455}. The search parameters were: exhaustiveness of $20$ and spacing of $1.00\,\AA$, for both. Furthermore, in this procedure, a second \emph{in-house} Python script was used to automatically run the virtual screening and select, from the results, the top 1000 compounds with the best binding energy for a second filtering using molecular docking.

\subsection{Molecular Docking}
Using the Autodock GPU software~\cite{Morris-J.Comput.Chem.-30-2785,Santos_Martins-ChemRxiv-2019}, a molecular docking was performed with the top 1000 molecules. This new Autodock uses the Solis-Wets algorithm. A new gridbox was prepared on the previously described binding sites, using a spacing of $0.375$ in both cases. The parameters of the molecular docking were: population size of $350$, number of evaluations of $2500000$, number of generations of $27000$, a mutation ratio of $0.02$, a crossover ratio of $0.8$ and number of run equal $100$. In each docking trial, the selection of the best poses was made based on the 2D score (2DS) described by Blanco 2019. It normalizes two variables of its own for each pose using the following arithmetic combination $2DS = N_{pop} - NMBE$, where $N_{pop}$ is the population of the cluster where the pose of the ligand was classified, and $NMBE$ is the average coupling energy. This procedure was carried out in the same way using another \emph{in-house} Python script, to automate the procedures and make a new selection of the top 10 best compounds based on binding energies, both for the RBD and for the cryptic site. Each selected pose was extracted with PyMol v2.3.5 and converted to PDB format for subsequent simulations.

\subsection{Data analysis and biomolecular graphics}
The resulting data from virtual screening was treated using a Python and C++ \emph{in-house} scripts. The first one was used to select the top of the best ligands based on the binding energy and the second one, to convert the binding energy ($\Delta G$) to the dissociation constant ($k_d$), through the equation $\Delta G = RTln(k_d)$ ($R$ is the gas constant and $T$ is the temperature, taken equal to $300\,K$).The generated poses that showed the best interaction energies were analyzed with PLIP~\cite{Salentin-NucleicAcidsRes.-43-443} to determine the type and number of interactions between compounds and receptor. The figures were rendered with VMD software~\cite{Humphrey-J.Mol.Graphics-14-33}.

\section{Results}
The untimely appearance of SARS-CoV-2, as a result of a kind of global crisis, has affected not only China, but hundreds of other countries, where the impact has been worse. Therefore, many research groups in the world are putting all their interest and efforts on it. The main interest is to block the virus from entering the host cell, and give the immune system time to recognize and eliminate it. For this purpose, multiple approaches are explored. The one we explored here was to identify potential small molecules with the ability to interact with key residues involved in the molecular recognition of the ACE2-RBD complex, which was successfully achieved, and it is shown in the following paragraphs of this section.

\subsection{Virtual screening and molecular docking}
\begin{figure}[htpb]
\centering
\includegraphics[width=120mm,keepaspectratio=true]{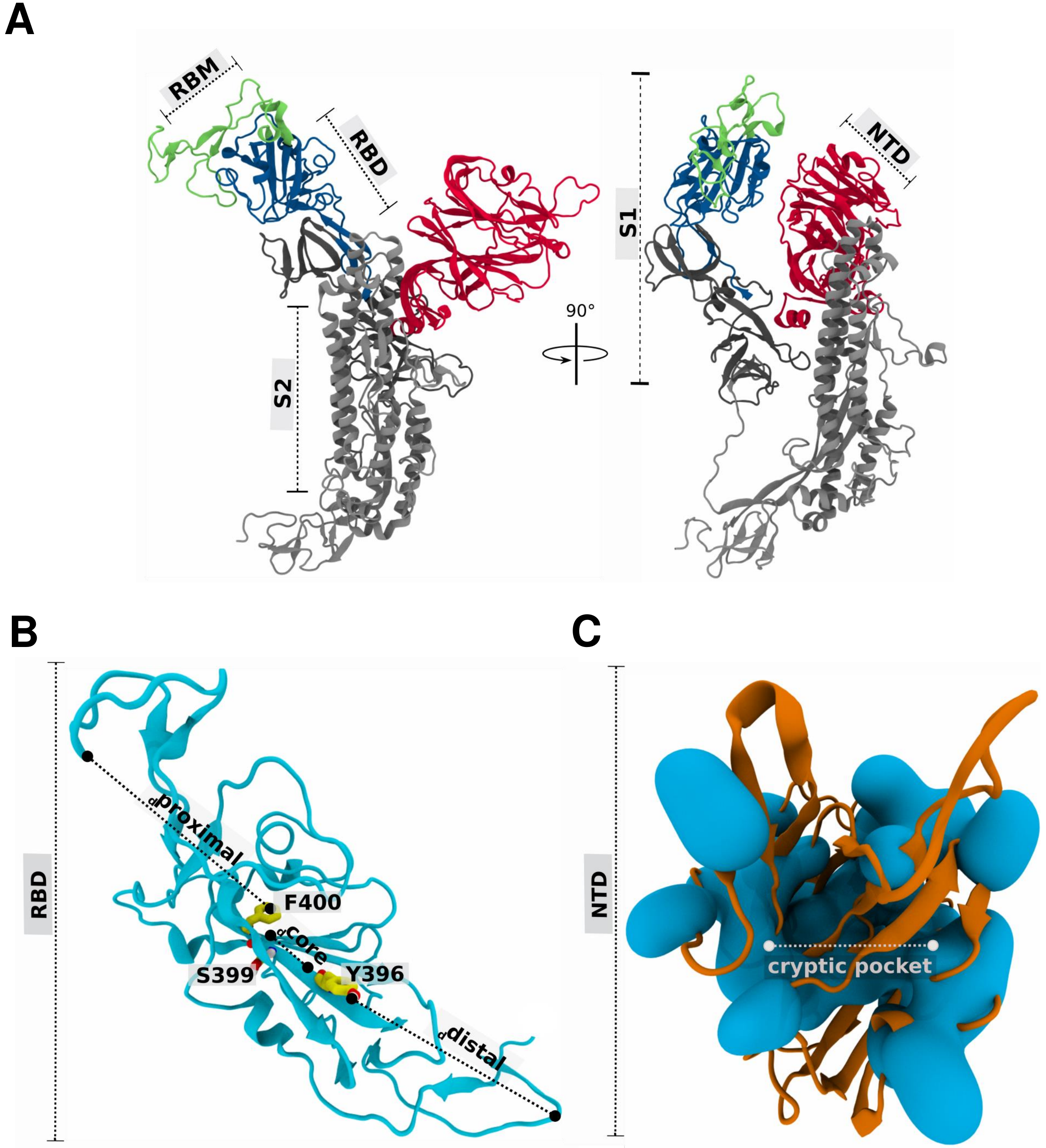}
\caption{Anatomy and binding site of Spike glycoprotein. (A) Front view of the monomeric protein S is shown, where the receptor-binding motif (RBM) is colored in green, receptor-binding domain (RBD) in blue and S2 domain in silver. Moreover, the lateral view (left) of the protein shows the S1 domain colored in dark-silver and N-terminal domain (NTD) in red. (B) The three referential residues Y396, S399, and F400 are represented in yellow-sticks. The proximal region, the core and distal region are signalized in dotted lines. (C) The cryptic pocket is shown in cartoon while the principal residues are shown in surface.}
\label{Fig:Fig_01}
\end{figure}

As reported in various studies, the Recognition Binding Domain of SARS-Cov-2 spike protein mediates the entry of the virus into the host cell through the human ACE2 receptor~\cite{Matsuyama-J.Virol.-84-12658,Glowacka-J.Virol.-85-4122,Shulla-J.Virol.-85-873,Song-PLOSPathogens-14-1007236,Kawase-J.Virol.-93-19,Letko-NatureMicrobiology-5-562,Monteil-Cell-181-7}. Therefore, this domain was considered the binding site for virtual screening evaluations, since the main amino acids involved in the interaction are found in this region (Y396, S399 and F400) (figure~\ref{Fig:Fig_01}A), and, which are also points of access to the distal and proximal regions of the domain that allowed us to cover most of the domain. On the other hand, we introduced the search for another region of the protein as an interaction site with small molecules, this site is called cryptic site, which was predicted with the Cryptosite server~\cite{Cimermancic-J.Mol.Biol.-428-709}, managing to observe this cryptic site with potential drugability, located in the N-Terminal Domain (figure~\ref{Fig:Fig_01}C).

In the virtual screening, the compounds were classified based on their free binding energy, $\Delta G$, obtaining the top 1000 compounds with the best binding and cryptic site affinity. Subsequently, the evaluation of the top 1000 compounds, by  molecular docking, allowed identifying the top 10 ligands that have high affinity with the RBD of S protein, in the binding site and cryptic site ($\Delta G \leq -8.53\, kcal/mol$ and $\Delta G \leq -10.66\, kcal/mol$, respectively).

To determine the molecular recognition and test that these interact with key residues, the analysis was performed with PLIP software, where it was observed that the ligands interacted forming hydrogen bonds, salt bridges, aromatic interactions and hydrophobic interactions (tables~\ref{Tab:Tab_01} and~\ref{Tab:Tab_02}). In the binding site, a frequently observed interaction, in eight of the best compounds studied, was that of hydrogen bonds, with at least one of the residues N487 and/or Y489 (table~\ref{Tab:Tab_01}), which are important for their participation in binding to ACE2. While at the cryptic site, hydrogen-bridged interactions were observed more frequently with residues H207, Q173, and R190 (table~\ref{Tab:Tab_02}).

\begin{figure}[htpb]
\centering
\includegraphics[width=120mm,keepaspectratio=true]{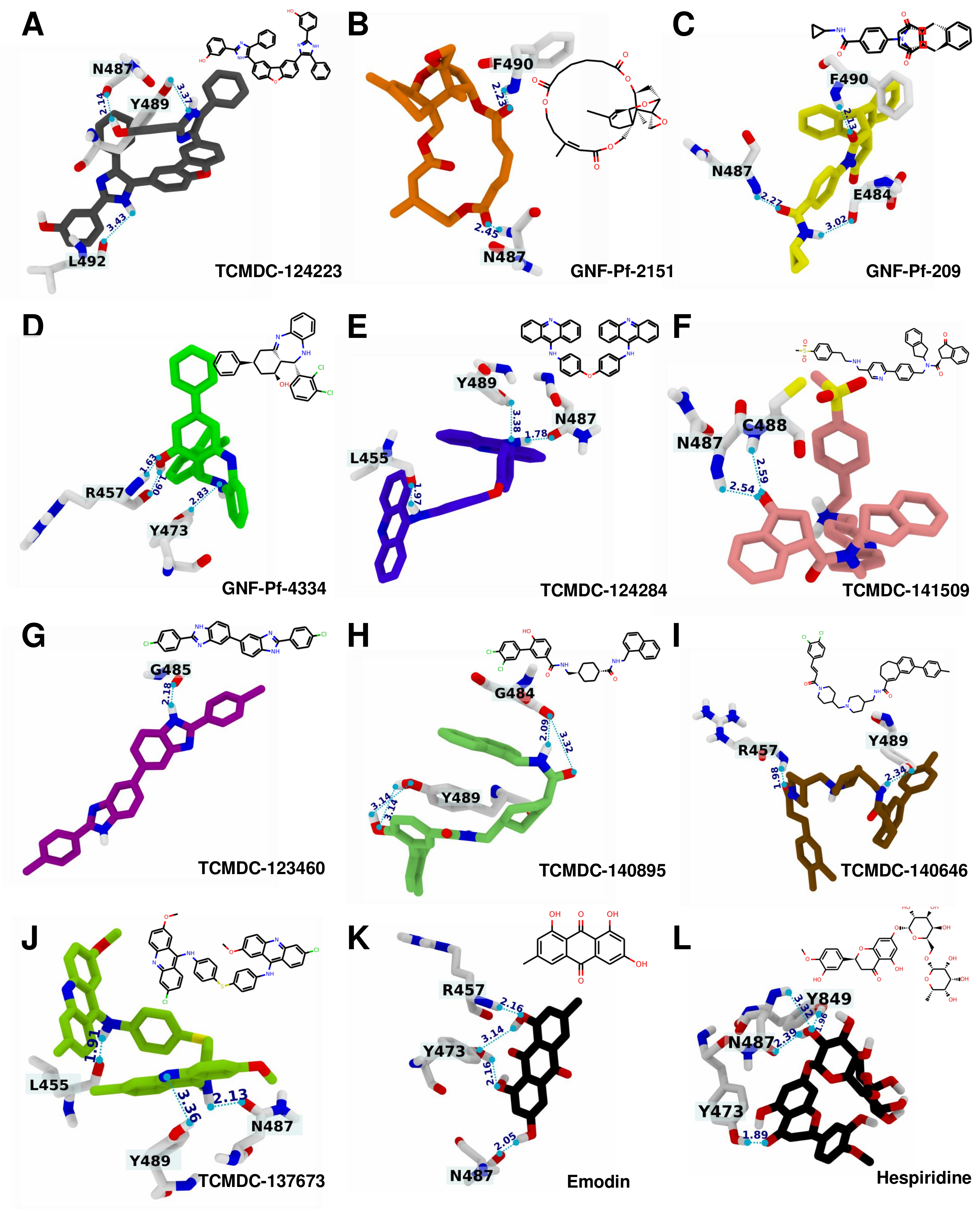}
\caption{Molecular recognition in RBD. The binding modes of the top 10 ligands obtained from the molecular docking assays,  and positive controls are represented in black-sticks. The receptor residues are represented in white-sticks, and the ligands are shown in different colors. The hydrogen bonds are shown in sky blue dotted lines.}
\label{Fig:Fig_02}
\end{figure}

Compound TCMDC-124223 presented the most favorable interaction energy ($\Delta G = -8.53\, kcal/mol$, $k_d = 530.53\, nM$) at the binding site (RBD) of S protein. This compound is molecularly recognized by the formation of hydrogen bonds with residues N487, Y489 and L492, at distances of $2.14\, \AA$, $3.37\, \AA$ and $3.43\, \AA$, respectively (figure~\ref{Fig:Fig_02}A). Furthermore, it presents an aromatic interaction with residue Y489 and hydrophobic interactions with residues F456, Y473, A475, Q484, Y489, F490 and L492 (table~\ref{Tab:Tab_01}).

\newpage
\newgeometry{left=2cm}
\begin{landscape}
\begin{table}[htbp]
\tiny
\caption{Ligand interactions with the Recognition Binding Domain of domain $S1$ (RBD-$S1$) Spike from SARS-CoV-2.}
\label{Tab:Tab_01}
\begin{center}
\begin{tabular}{|c|c|c|cccc|cc|cc|}
  \hline \hline
  \multirow{3}{2cm}{Ligand} &  &  & \multicolumn{8}{c|}{Interactions} \\
                        & $\Delta G$  & $k_d$ & \multicolumn{4}{c|}{H-Bond} & \multicolumn{2}{c|}{Aromatic} & \multicolumn{2}{c|} {Hydrophobic} \\
                         & &  &  Number & Residue & Functional group & Distance & Number & Residue & Number & Residue \\
  \hline
  TCMDC-124223 & $-8.53$ & $530.53$ & $3$  & Asn487(O-OH) & hydroxyl          & $2.14$ & $1$ & Tyr489 & $7$ & Phe456,Tyr473,Ala475, \\
               &         &          &      & Tyr489(OH-N) & secondary amino   & $3.37$ &     &        &     & Glu484, Tyr489, Phe490, \\
               &         &          &      & Leu492(O-HN) & secondary amino   & $3.43$ &     &        &     & Leu492                     \\
  \hline
  GNF-Pf-2151  & $-8.49$ & $567.73$ & $2$ & Asn487(NH-O)   & ester             & $2.45$ &     &        & $3$ & Phe456, Glu484, Tyr489   \\
               &         &          &     & Phe490(NH-O)   & ester             & $2.23$ &     &        &     &                       \\
  \hline
  GNF-Pf-209   & $-8.41$ & $650.12$ & $3$ & Glu484(O-HN)   & carboxamide       & $3.02$ & $1$ & Tyr489 & $5$ & Phe456, GLu484, Phe486, \\
               &         &          &     & Asn487(NH-O)   & carboxamide       & $2.27$ &     &        &     & Tyr489, Phe490 \\
               &         &          &     & Phe490(NH-O)   & secondary amino   & $2.13$ &     &        &     &                       \\
  \hline
  GNF-Pf-4334  & $-8.35$ & $719.67$ & $3$ & Arg457(NH-O)   & hydroxyl          & $1.63$ & $2$ & Tyr421,& $3$ & Lys417, Phe456, Ala475 \\
               &         &          &     & Arg457(O-HO)   & hydroxyl          & $1.90$ &     & Tyr473 &     &    \\
               &         &          &     & Tyr473(OH-N)   & secondary amino   & $2.83$ &     &        &     &                       \\
  \hline
  TCMDC-124284 & $-8.31$ & $770.13$ & $3$ & Leu455(O-HN)   & secondary amino   & $1.97$ &     &        & $5$ & Lys417, Tyr421, Leu455, \\
               &         &          &     & Asn487(O-HN)   & secondary amino   & $1.78$ &     &        &     & Phe456, Tyr489 \\
               &         &          &     & Tyr489(OH-N)   & secondary amino   & $3.38$ &     &        &     &            \\
  \hline
  TCMDC-141509 & $-8.29$ & $796.66$ & $2$ & Asn487(NH-O)   & ketone            & $2.54$ & $2$ & Phe456,& $5$ & Glu484, Phe486, Tyr489, \\
               &         &          &     & Cys488(NH-O)   & ketone            & $2.59$ &     & Phe490 &     & Phe490, Gln493          \\
  \hline
  TCMDC-123460 & $-8.24$ & $867.08$ & $1$ & Gly485(O-HN)   & secondary amino   & $2.18$ &     &        & $6$ & Glu484, Phe486, Tyr489,  \\
               &         &          &     &                &                   &        &     &        &     & Phe490, Leu492, Gln493   \\
  \hline
  TCMDC-140895 & $-8.22$ & $896.96$ & $4$ & Glu484(OH-O)    & carboxamide      & $3.32$ & $2$ & Phe456,& $4$ & Phe456, Glu484, Tyr489,   \\
               &         &          &     & Glu484(O-HN)    & secondary amino  & $2.09$ &     & Tyr489 &     & Phe490    \\
               &         &          &     & Tyr489(OH-O)    & hydroxyl         & $3.14$ &     &        &     & \\
               &         &          &     & Tyr489(O-HO)    & hydroxyl         & $3.14$ &     &        &     &                       \\
  \hline
  TCMDC-140646 & $-8.18$ & $959.84$ & $2$ & Arg457(NH-O)    & carboxamide      & $1.98$ & $1$ & Phe456 & $5$ & Lys417, Leu455, Phe456, \\
               &         &          &     & Tyr489(O-HN)    & secondary amino  & $2.34$ &     &        &     & Tyr489, Phe490 \\
  \hline
  TCMDC-137673 & $-8.15$ & $1009.9$ & $3$ & Leu455(O-HN)    & secondary amino  & $1.91$ & $1$ & Phe486 & $5$ & Lys417, Tyr421, Leu455,  \\
               &         &          &     & Asn487(O-HN)    & secondary amino  & $2.13$ &     &        &     & Phe456, Tyr489 \\
               &         &          &     & Tyr489(OH-N)    & tertiary amino   & $3.36$ &     &        &     &   \\
  \hline
  \hline
\end{tabular}
\begin{flushleft}
\tiny {Note: $\Delta G$ is in units of kcal/mol. $k_d$ is in units of nM and distances are in $\AA$.}
\end{flushleft}
\end{center}
\end{table}
\end{landscape}
\restoregeometry

\begin{figure}[htpb]
\centering
\includegraphics[width=120mm,keepaspectratio=true]{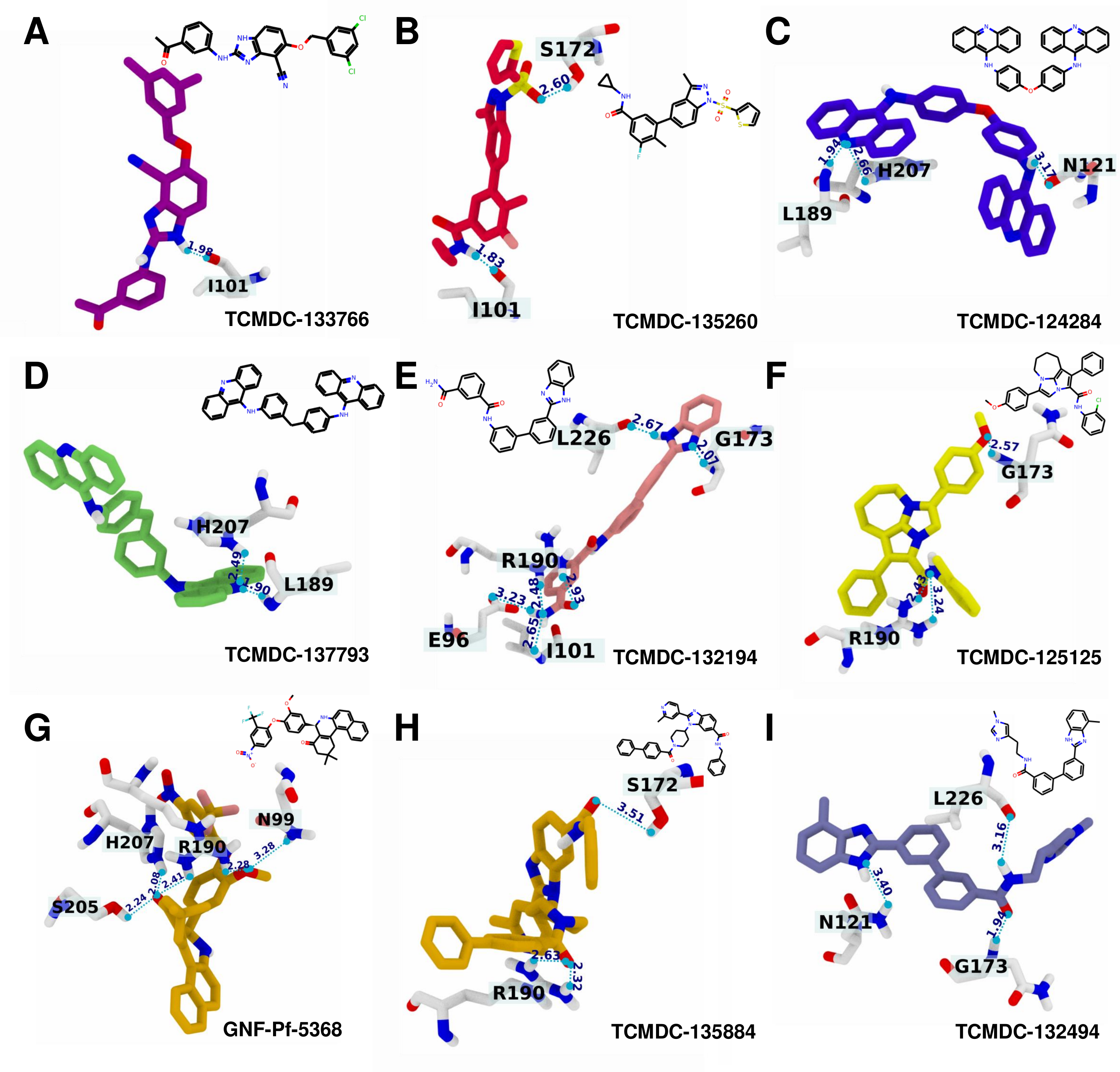}
\caption{Interactions in the cryptic site. The binding modes obtained from molecular docking assays are represented in sticks. The receptor residues are represented in white-sticks, and the ligands are shown in different colors.}
\label{Fig:Fig_03}
\end{figure}

At the cryptic site, compound TCMDC-133766 presented the best interaction energy ($\Delta G = -10.66\, kcal/mol$, $k_d = 14.37\, nM$), which exceeds the compounds evaluated in the binding site. The mode of binding of this compound is mediated by the formation of a hydrogen bridge with residue I101, at a distance of $1.98\, \AA$. Likewise, in the formation of the complex, aromatic interactions with residues F92, T104, F192 and R190 and hydrophobic interactions with residues F92, I101, T104, F106, V126, F175, F192, F194, L226 and Y240 (Figure~\ref{Fig:Fig_03}A, and table~\ref{Tab:Tab_02}).

\newpage
\newgeometry{left=2cm}
\begin{landscape}
\begin{table}[htbp]
\tiny
\caption{Ligand interactions in the predicted Spike cryptic pocket from SARS-CoV-2.}
\label{Tab:Tab_02}
\begin{center}
\begin{tabular}{|c|c|c|cccc|cc|cc|}
  \hline \hline
  \multirow{3}{2cm}{Ligand} &  &  & \multicolumn{8}{c|}{Interactions} \\
                        & $\Delta G$  & $k_d$ & \multicolumn{4}{c|}{H-Bond} & \multicolumn{2}{c|}{Aromatic} & \multicolumn{2}{c|} {Hydrophobic} \\
                         & &  &  Number & Residue & Functional group & Distance & Number & Residue & Number & Residue \\
  \hline
   TCMDC-133766 & $-10.66$  & $14.37$  & $1$ & Ile101(NH-O)  & secondary amino & $1.98$  & $4$ & Phe92, Trp104,  & $10$ & Phe92, Ile101, Trp104, Phe106, \\
                &           &          &     &               &                 &         &     & Phe192, Arg190  &      & Val126, Phe175, Phe192, Phe194, \\
                &           &          &     &               &                 &         &     &                 &      & Leu226, Thr240 \\
  \hline
   TCMDC-135260 & $-10.24$  & $29.28$  & $2$ & Ile101(OH-N) & secondary amino  & $1.83$  &     &                 & $10$ & Phe92, Ile101, Trp104, Ile119, \\
                &           &          &     & Ser172(O-HO) & sulfone          & $2.60$  &     &                 &      & Phe175, Phe192, Phe194, Ile203, \\
                &           &          &     &              &                  &         &     &                 &      & Leu226, Val227 \\
  \hline
   TCMDC-124284 & $-10.21$  & $30.81$  & $3$ & Asn121(O-HN)  & secondary amino & $3.17$  & $1$ & Phe192 & $10$ & Phe92, Ile101, Trp104, Val126, \\
                &           &          &     & Leu189(NH-N)  & tertiary amino  & $1.94$  &     &        &      & Gln173, Phe175, Phe192, Ile203, \\
                &           &          &     & His207(NH-N)  & tertiary amino  & $2.66$  &     &        &      & Leu226, Val227 \\
  \hline
   TCMDC-135232 & $-10.14$  & $34.69$  &     &               &                 &         & $2$ & Phe92, & $8$  & Ile101, Trp104, Ile119, Val126, \\
                &           &          &     &               &                 &         &     & Arg190 &      & Gln173, Phe175, Phe192, Leu226 \\
  \hline
   TCMDC-137793 & $-10.09$  & $37.76$  & $2$ & Leu189(NH-N) & tertiary amino   & $1.90$  & $2$ & Arg190, & $9$ & Glu96, Trp104, Val126, Phe175, \\
                &           &          &     & His207(NH-N) & tertiary amino   & $2.49$  &     & His207  &     & Asn188, Arg190, Ile203, His207, \\
                &           &          &     &              &                  &         &     &         &     & Val227 \\
  \hline
  TCMDC-132194  & $-10.94$  & $41.10$  & $6$ & Glu96(O-HN)  & primary amino    & $3.23$ &  &  & $9$  & Ile101, Trp104, Ile119, \\
                &           &          &     & Ile101(NH-N) & primary amino    & $2.65$ &  &  &      & Val126, Gln173, Phe175, \\
                &           &          &     & Gln173(NH-N) & tertiary amino   & $2.07$ &  &  &      & Phe192, Leu226, Val227 \\
                &           &          &     & Arg190(NH-N) & primary amino    & $2.48$ &  &  &      &  \\
                &           &          &     & Arg190(NH-O) & carboxamine      & $2.93$ &  &  &      &  \\
                &           &          &     & Leu226(O-HN) & secondary amino  & $2.67$ &  &  &      &  \\
  \hline
  TCMDC-125125  & $-9.96$   & $47.06$  & $3$ & Gln173(NH-O) & ether            & $2.57$ & $1$ & Arg190  & $10$  & Ile101, Trp104, Ile119, Val126, \\
                &           &          &     & Arg190(NH-N) & secondary amino  & $3.24$ &  &  &  & Gln173, Phe175, Phe192, Ile203,  \\
                &           &          &     & Arg190(NH-N) & secondary amino  & $2.43$ &  &  &  & Leu226, Val227 \\
  \hline
   GNF-Pf-5368  & $-9.93$   & $49.51$  & $5$ & Asn99(NH-O)   & ether  & $3.28$ & $2$ & Phe175, & $6$ & Gln173, Phe175, Asn188, Phe192, \\
                &           &          &     & Arg190(NH-O)  & ketone & $2.41$        &     & His207  &     & His207, Leu226 \\
                &           &          &     & Arg190(NH-O)  & ether  & $2.28$ &  &  &  &  \\
                &           &          &     & Ser205(OH-O)  & ketone & $2.24$ &  &  &  &  \\
                &           &          &     & His207(NH-O)  & ketone & $2.08$ &  &  &  &  \\
  \hline
  TCMDC-135884  & $-9.91$   & $51.22$  & $3$ & Ser172(OH-O)  & carboxamine & $3.51$ & $1$ & Phe192  & $10$ & Phe92, Ile101, Trp104, Val126, \\
                &           &          &     & Arg190(NH-N)  & carboxamine & $2.32$ &  &  &  & Gln173, Phe175, Phe192, Ile203, \\
                &           &          &     & Arg190(NH-N)  & carboxamine & $2.63$ &  &  &  & Leu226, Val227 \\
  \hline
  TCMDC-132494  & $-9.84$   & $57.67$  & $3$ & Asn121(NH-N) & secondary amino & $3.40$ & $1$ & Arg190 & $7$  & Arg102, Trp104, Val126, Phe175, \\
                &           &          &     & Gln173(NH-O) & carboxamine     & $1.94$ &  &  &  & Phe192, Leu226, Val227 \\
                &           &          &     & Leu226(O-HN) & secondary amino & $3.16$ &  &  &  &  \\
  \hline
  \hline
\end{tabular}
\begin{flushleft}
\tiny {Note: $\Delta G$ is in units of kcal/mol. $k_d$ is in units of nM and distances are in $\AA$.}
\end{flushleft}
\end{center}
\end{table}
\end{landscape}
\restoregeometry

Additionally, two drugs (Hespiridine and Emodin), approved by the FDA, were taken as controls for the analysis of molecular docking. Emodin presented a binding energy $-5.62\, kcal/mol$; while, the compound Hespidiridine, presented an energy of $-2.46\, kcal/mol$ (table~\ref{Tab:Tab_03}). Similarly, analysis with PLIP software found that these drugs form hydrogen bridge-type interactions with important residues, such as Y473 and N487, and hydrophobic interactions (figure~\ref{Fig:Fig_02}K).

\newpage
\newgeometry{left=2cm}
\begin{landscape}
\begin{table}[htbp]
\tiny
\caption{Ligand-control interactions with the Recognition Binding Domain of domain $S1$ (RBD-S1) Spike from SARS-CoV-2.}
\label{Tab:Tab_03}
\begin{center}
\begin{tabular}{|c|c|ccc|cc|cc|}
  \hline \hline
  \multirow{3}{2cm}{Ligand} &  &  & \multicolumn{6}{c|}{Interactions} \\
                        & $\Delta G$  & \multicolumn{3}{c|}{H-Bond} & \multicolumn{2}{c|}{Aromatic} & \multicolumn{2}{c|} {Hydrophobic} \\
                        &             &  Number & Residue  & Distance & Number & Residue & Number & Residue \\
  \hline
  Emodin      & $-5.62$  & $4$ & Arg457(NH-O)  & $2.16$  &  & & $4$ & Tyr421, Leu455,  \\
              &          &     & Tyr473(O-HO) & $3.14$  &  & &     & Tyr473, Ala475   \\
              &          &     & Tyr473(OH-O) & $2.16$  &  & &     &  \\
              &          &     & Asn487(O-HO) & $2.05$  &  & &     &  \\
  \hline
  Hespiridine & $-2.46$  & $4$ & Tyr473(OH-O) & $1.89$  &  & &  $4$   & Lys417, Tyr421,  \\
              &          &     & Asn487(NH-O) & $3.32$  &  & &     & Leu455, Phe456 \\
              &          &     & Asn487(O-HO) & $2.39$  &  & &     &  \\
              &          &     & Tyr489(OH-O) & $1.96$  &  & &     &  \\
  \hline
  \hline
\end{tabular}
\begin{flushleft}
\tiny {Note: $\Delta G$ is in units of kcal/mol. $k_d$ is in units of nM and distances are in $\AA$.}
\end{flushleft}
\end{center}
\end{table}
\end{landscape}
\restoregeometry

\section{Discussion}
The current SARS-CoV-2 pandemic represents one of the greatest challenges due to its easy transmission~\cite{Yang-IntegrMed.Res.-9-100426}. Despite various investigations being conducted, so far no treatment is effective in fighting this virus. Computational approaches are promising alternatives for finding potential inhibitors through drug repositioning. Therefore, this research focused on the search for RBD inhibitors for S protein that can prevent recognition by the ACE2 receptor.

The results of the virtual screening of $13,102$ compounds allowed the identification of potential inhibitors that can be used in the treatment of Covid-19, including TCMDC-124223, GNF-Pf-2151 and GNF-Pf-209 for the binding site (RBD) and TCMDC-133766, TCMDC-135260 and TCMDC-124284 for the cryptic site (NTD). These compounds have previously been used as antimalarial potentials~\cite{Voorhis-PLOSPathogens-12-1005763}, which demonstrates their versatility.

Various studies indicate that the amino acids located in the RBM contribute substantially to the recognition and subsequent entry of the virus into the cell~\cite{Wang-Cell-181-9,Yi-CellMol.Immunol.-17-621,Liu-PLOSPathogens-16-1008421}. Our analyzes demonstrate that six compounds manage to interact by forming hydrogen bonds with residue N487 (table~\ref{Tab:Tab_01}), and polar contacts with residue Y83 of the ACE2 receptor~\cite{Stawiski-bioRxiv-2020}. Likewise, eight compounds show hydrophobic interactions with the Y489 residue (table~\ref{Tab:Tab_01}). This type of interaction with the K31 residue of ACE2 has been shown to be conserved among variants of this protein~\cite{Hussain-J.Med.Virol.-1-7}. This suggests that the interactions formed could cause modifications in the contact surface, preventing recognition.

Furthermore, the formation of hydrophobic interactions with the K417 residue was observed in at least four compounds in the top 10. This residue is known to be located in the loop of the proximal part of the RBM and has a high-energy  contribution ($-93.13\, kcal/mol$) during molecular recognition with ACE2~\cite{ArmijosJaramillo-Evol.Appl.-00-1}. In this sense, those compounds that present some form of favorable interaction with this residue could promote a competitive type inhibition on the contact surface. However, chemical modifications could be introduced, to raise the binding energy. Similar studies have highlighted the importance of this residue in the inhibition of RBD~\cite{Li-Science-309-1864,Lan-Nature-581-215}.

As other studies point out, the drugability of cryptic sites allows modifying the thermodynamic or structural characteristics of proteins~\cite{Cimermancic-J.Mol.Biol.-428-709}. In this study, we report a potential drugable cryptic site, located in the N-Terminal domain of the Spike protein. The structure that makes up this cryptic site allows the virus to merge with the cell membrane~\cite{Li-Science-309-1864,Bertram-J.Virol.-85-13363}, so its inhibition could indirectly prevent the entry of the virus. The drug that presented the best mode of coupling on this cryptic site (TCMDC-133766) has shown a higher affinity ($-10.66\, kcal/mol$) compared to the best compounds tested in the RBD. In this sense, it is possible to evaluate the synergism of the inhibition of RBD and NTD, in such a way that the activity of the Spike protein can be suppressed with greater effectiveness.

On the other hand, it was observed that the compounds with the best interaction energy have common functional chemical groups of other inhibitor molecules~\cite{Robertson-Curr.Opin.Struct.Biol.-17-674}. The functional groups of the top compounds in the binding site (RBD) are mainly of the secondary amino, hydroxyl, and carboxamine type. In the cryptic site, the most representative functional groups are the secondary amines, tertiary amino and carboxamines (tables~\ref{Tab:Tab_01} and~\ref{Tab:Tab_02}). These functional groups, which form hydrogen bonds, can be considered for rational drug design based on a pharmacophore model, as has been proposed for other drugable targets~\cite{Enriquez-1996}.

For the control of molecular docking evaluations, two compounds were used, Emodine and Hespiridine, which are approved by the FDA~\cite{Ho-AntiviralRes.-74-92,Coppola-Med.Hypotheses-140-109766}. These drugs interact with important residues in the RBD (N487 and Y489). However, it has been observed that these compounds present lower interaction energies ($-5.62\, kcal/mol$ and $-2.46\, kcal/mol$ respectively) compared to the top of the evaluated compounds. This suggests that the compounds reported in this work could improve the inhibitory activity on the Spike protein.

Finally, our studies suggest a pharmacological potential of the present molecules against the Spike protein of SARS-CoV-2. \emph{In-vivo} studies can confirm the inhibitory activity of these compounds. Furthermore, the functional groups of these drugs can be used to search for similar compounds in different databases.

\section{Conclusion}
Our approach has allowed us to identify a set of small molecules with the capacity to interrupt the interaction of ACE2-RBD, since their molecular recognition is associated with key residues in the interaction of the ACE2-RBD complex, and that these can also interact with the cryptic site and can reduce the interaction of this complex, through synergism. In therapy, that it is called \emph{combination therapy}, which has shown better benefits in the treatment of the disease. However, these compounds must be tested \emph{in-vitro} to demonstrate their activity, and undergo the respective clinical tests. It is worth mentioning that the functional groups of these compounds could be used to perform pharmacophore models, and to identify FDA-approved drugs and speed up clinical tests.

\section{Acknowledgments}
Part of the results presented here were developed with the help of CENAPAD-SP (Centro Nacional de Processamento de Alto Desempenho em S\~{a}o Paulo) grant UNICAMP/FINEP-MCT and CENAPAD-UFC (Centro Nacional de Processamento de Alto Desempenho, at Universidade Federal do Cear\'a).

\newpage
\bibliographystyle{elsarticle-num}
\bibliography{Geordano}

\end{document}